\documentclass[12pt]{iopart}
\usepackage{iopams}  
\usepackage{graphicx}
\usepackage{titlesec}
\usepackage{dcolumn}
\usepackage{appendix}
\usepackage{xcolor}
\usepackage{caption}
\usepackage{geometry}
\begin{document}

\title{Formation mechanism of the rotating spoke in partially magnetized plasmas}

\author{Liang Xu,$^1$ Denis Eremin,$^1$ Andrei Smolyakov,$^2$ Dennis Kr\"{u}ger,$^1$ Kevin K\"{o}hn,$^1$ and Ralf Peter Brinkmann$^1$}

\address{$^{1)}$Institute for Theoretical Electrical Engineering, Ruhr-University Bochum, D-44780 Bochum, Germany}
\address{$^{2)}$Department of Physics and Engineering Physics, University of Saskatchewan, Saskatoon, Saskatchewan S7N 5E2, Canada}
%\ead{submissions@iop.org}
%\vspace{10pt}
%\begin{indented}
%\item[]August 2017
%\end{indented}
\begin{abstract}
Rotating spokes commonly occur in partially magnetized plasmas devices. In this paper, the  driving mechanism behind the formation of an m=1 rotating spoke mode in a magnetically enhanced hollow cathode arc discharge is investigated by means of 2D radial-azimuthal particle-in-cell/Monte Carlo collision simulations with a uniform axial magnetic field. 
We find that the formation of the spoke potential hump region can be explained as a result of the positive anode sheath collapse due to the lower hybrid type instability evolving into the long wavelength regime. It is shown that an initial short-wavelength instability in the non-neutral anode sheath undergoes a sequence of transitions into the large scale mode. The sheath non-neutrality effect on the instability is considered and incorporated in the two-fluid linear theory of the lower hybrid instability. The unstable modes predicted by the theory in the linear phase and nonlinear evolution are in good agreement with the fluctuation modes developed in the particle simulations.

\end{abstract}

\newpagestyle{main}{            
    \setfoot{}{}{\thepage} 
%    \setfoot{}{}{}      
%    \headrule                                    
%    \footrule                                    
}
\pagestyle{main}    %使用该style

%
% Uncomment for keywords
%\vspace{2pc}
%\noindent{\it Keywords}: XXXXXX, YYYYYYYY, ZZZZZZZZZ
%
% Uncomment for Submitted to journal title message
%\submitto{\JPA}
%
% Uncomment if a separate title page is required
%\maketitle
% 
% For two-column output uncomment the next line and choose [10pt] rather than [12pt] in the \documentclass declaration
%\ioptwocol
%\twocolumn

\section{Introduction}

Partially magnetized plasmas, using external magnetic field $\mathbf{B}$ perpendicular to applied electric field $\mathbf{E}$ to confine electrons and generate high plasma density in low pressures, represent an advanced plasma source in technology and industry, such as magnetrons, Hall effect thrusters and magnetically enhanced glow/arc discharges \cite{Anders2017,Boeuf2017,brinkmann2015physics,fietzke2009magnetically,gudmundsson2020physics}. These discharges feature magnetized electrons subject to the $\mathbf{E} \times \mathbf{B}$ confinement (electron Larmor radius $\rho_e$ is smaller than the plasma size $L$, $\rho_e<L$) and non- or weakly magnetized ions (ion Larmor radius $\rho_i\gtrsim L$), which are accelerated almost collisionlessly in the applied electric field  for the application purpose. Inherently, due to the imposed magnetic field, various oscillation modes from high frequency (on the order of ${\rm MHz}$) \cite{forslund1971nonlinear,janhunen2018evolution,lafleur2016theory,adam2004study,lundin2008anomalous,mcbride1972theory,tsikata2014axially,charoy2021interaction} to low frequency (on the order of ${\rm KHz}$) \cite{hara2014mode,barral2009low,romadanov2018hall,Kaganovich2020,panjan2017plasma,Hecimovic2018,hnilica2018,Ellison2012} can be excited and change the dynamics of electrons and heavy particles, which have been predicted by theory and reported in a large number of experiments and numerical simulations. In the low frequency regime, rotating spoke is one of the most prominent oscillation modes. Experiments demonstrated that rotating spokes are regions where the electric potential is locally enhanced (potential humps) and electrons are energized in a double layer surrounding the potential hump
region \cite{Anders2013,panjan2017plasma,panjan2014asymmetric}. The spokes and associated electron transport and heating have attracted  significant interests recently  \cite{BoeufPRL2013,MatyasPSST2019s,MazouffrePSST2019,koshkarov2019self, sengupta2021restructuring,kawashima2018numerical,Kaganovich2020,Rudolph2021,Ellison2012,lucken2019instability}. 
Unfortunately, the rotating spoke characteristics (rotation velocity, mode number, potential hump etc.) and its underlying physics are not well understood for a variety of possible configurations, thereby prohibiting the establishment of predictive model for $\mathbf{E}\times \mathbf{B}$ plasmas based applications. In this paper, we address the driving mechanism behind the formation of rotating spokes in a magnetized plasma source with externally applied electric field. 

We study the rotating spoke under conditions of magnetically enhanced hollow cathode arc discharge (ME-HCAD), a promising plasma source for large area film deposition \cite{fietzke2009magnetically,fietzke2010plasma,zimmermann2011spatially}. The plasma device consists of a hollow cathode with a coaxial anode outside, enclosing by a solenoid to generate a uniform axial magnetic field inside the hollow cathode and an expanding magnetic field in the plume. ME-HCAD produces plasma with extremely high electron density ($10^{18}-10^{20}{\rm m^{-3}}$) and large area plasma plume (up to $1 {\rm m^{2}}$) for film deposition. In our previous work using a cylindrical PIC/MCC approach \cite{Xu2021}, we reported the observation of a well-established and robust $m=1$ spoke mode in ME-HCAD. At the early linear phase of the numerical simulations, the unstable modes were identified as the lower hybrid type gradient drift instability under different magnetic fields. The short wavelength modes cascading to long wavelength modes were also seen after the instability saturation. However, the mechanism behind the nonlinear transition and the formation of the spoke in these conditions were not elaborated, which are the main topics in the present paper. 

In Ref. \cite{boeuf2019micro}, the nonlinear evolution of micro instabilities and the formation of macroscopic structures in the planar magnetron was studied using the axial-azimuthal 2D PIC/MCC method assuming  a uniform radial magnetic field and given ionization. The author of Ref. \cite{boeuf2019micro} firstly computed the axial dimension using 1D PIC/MCC with the same parameters of 2D model and the obtained electron density and ion density were loaded in the 2D model as the initial condition. The 2D simulations showed that the gradient drift instability is developed in the near anode region and the instability mode depends on the magnetic field strength. In later work, Boeuf and Takahashi \cite{Boeuf2020new} studied the formation of spoke in micro planar magnetron with a nonuniform radial magnetic field and found that the formation of spoke results from the Simon-Hoh instability evolving into the ionization instability which is facilitated by the electron heating due to inhomogenous magnetic field. In the present paper, we show in the conditions of magnetized discharge with the applied electric field and uniform magnetic field, the formation of spoke is the consequence of the positive anode sheath collapse resulting from the lower hybrid type instability modified by non-neutrality effects.
 
The spokes formation and nonlinear transitions in a cylindrical magnetron were investigated by the radial-azimuthal 2D PIC/MCC method taking into account the ionization and self-consistent discharge organization  from neutral background \cite{sengupta2021restructuring}. It was shown that an m=2 rotating spoke is triggered by the lower hybrid gradient drift instability and its dynamics is controlled by the evolution of radial electric field at the near cathode region. Effects of varying neutral pressure were also demonstrated.

In this paper, compared to the works in Refs. \cite{boeuf2019micro,sengupta2021restructuring}, we implement a different initial condition: a uniform plasma background, i.e., $n_e(x,y)=n_i(x,y)$ and $\triangledown_{x} n_e(x,y)=\triangledown_{y} n_e(x,y)=0$. By virtue of the initial condition, the instability evolution shows smooth transition from the instability excitement and saturation to the spoke formation. Our model clearly demonstrate  the onset of the  instability,  formation of the spoke potential, and  nonlinear development. Supporting theoretical analysis  provides physical insights on the spoke mechanism and the evolution of the electron density and electric field being the result of the instability transition in the non-neutral anode sheath. The paper is structured as follows. Section 2 outlines the numerical model and presents the modified theory of the gradient drift instability by incorporating the non-neutrality effect. Section 3 presents the main results and discussions on the spoke nonlinear dynamics. The work will be summarized in Section 4.

\section{The numerical model and linear theory of the gradient drift instability}

\begin{figure}
\center
\includegraphics[clip,width=0.33\linewidth]{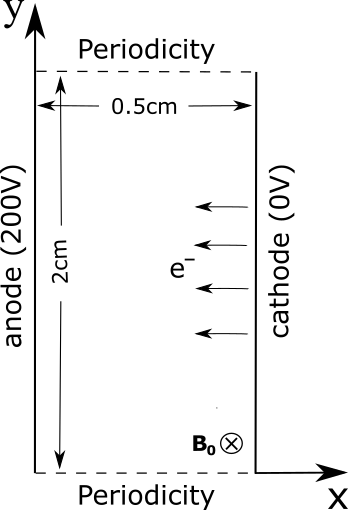}
\caption{2D radial (x)-azimuthal (y) PIC/MCC model of ME-HCAD. The working gas is argon and the gas pressure is $P=10 {\rm Pa}$ and the uniform axial magnetic field directs inward with the strength $B_0=[20 {\rm mT}, 80 {\rm mT}]$.}
\end{figure} 

\begin{table*}[]
    \centering
\begin{tabular}{p{4cm}|p{3cm}}
 \hline
 \hline
 Parameters & Value  \\
 \hline
 
Radial domain length $L_x$   & $5{\rm mm}$   \\
Azimuthal domain length $L_{y}$   & $20{\rm mm}$  \\
Anode voltage $U_a$  & $200{\rm V}$   \\
Cathode voltage $U_c$ & $0 {\rm V}$     \\
Gas pressure $P$   & $10{\rm Pa}$   \\
Gas temperature $T_g$ & $300{\rm K}$ \\
Axial magnetic field $B_0$ & $20-80 {\rm mT}$ \\
Initial plasma density $n_{0}$ & $2\times 10^{16}{\rm m^{-3}}$ \\
Initial electron temperature $T_{e0}$ & $ 5 {\rm eV}$ \\
Initial ion temperature $T_{i0}$ & $ 300{\rm K}$\\
Electron emission current density $j_{emit}$ & $1 {\rm A/m^2}$ \\
Thermionic electron temperature $T_{emit}$ & $2{\rm eV}$\\
Number of grid points &  $512 (y) \times 128 (x)$ \\
Cell size $\Delta x =\Delta y$ & $39{\rm \mu m}$\\
Time step $\Delta t$ & 16{\rm ps} \\
Number of particles per cell $N_{ppc}$ & 400 \\
 \hline
 \hline
\end{tabular}
\hspace*{-1cm}\caption{The table lists the physical and numerical parameters used in the PIC simulations.}
\end{table*}

To conduct the simulations of ME-HCAD, we use 2D-EDIPIC, an electrostatic explicit 2d3v PIC/MCC code, which was benchmarked against many other codes \cite{charoy2019,villafana2021}. The radial ($x$) and azimuthal ($y$) dimensions are resolved. We map the cylindrical geometry of ME-HCAD onto a Cartesian coordinate system under the assumption that the curvature effect does not play a significant role. In the Cartesian system, the electric field ${\bf E_0}$ is oriented along the discharge axis (x-direction, positive from anode to cathode), and a magnetic field ${\bf B_0}$ is taken along the z-direction (axially inward). Electrons drift azimuthally in the $y$ ($\pm {\bf E_0}\times {\bf B_0}$) direction. The model schematic is illustrated in Fig. 1. The discharge parameters we adopted are from the ME-HCAD experiments \cite{fietzke2010plasma} and listed in Table 1 together with the numerical parameters.

The initial state of the simulation is a stationary Maxwellian distribution for both electrons and ions with a homogeneous plasma background. The magnetic field strength is in the range of $B_0=20-80 {\rm mT}$, giving magnetized electrons and non-magnetized ions. Also, the magnetic field strength results in the positive anode sheath, i.e., the plasma potential is lower than the anode potential, which is of significant importance for the instability excitement and spoke formation in our cases. The simulation box area is $L_x \times L_y=5 {\rm mm} \times 20 {\rm mm}$. The cathode is grounded, the applied anode voltage is $U_a=200 {\rm V}$ and the azimuthal boundaries are periodical. The wave vector in the azimuthal direction is constrained by periodicity of the simulation domain to $k_y L_y = 2\pi m$. For particle boundaries, ions are absorbed at both electrodes; electrons are absorbed at the anode but specularly reflected at the cathode. The working gas is argon, the gas pressure is $P=10{\rm Pa}$ and the neutral depletion is not considered. The elastic, excitation and ionization electron-neutral collisions and charge exchange ion-neutral collision are implemented. The electron-neutral collision cross sections are those of Phelps \cite{phelps1999cold} and the charge exchange cross section is set to be $5.53\times10^{-19} {\rm m^{2}}$. 

We note that, to lower the computation cost of executing 2D-EDIPIC, the cathode thermionic electron emission is artificially reduced to keep the plasma density in the range of $10^{15}-{10^{17}} {\rm m^{-3}}$, which is much smaller compared to the experimentally observed values. This assumption can introduce the Debye length effect on the dynamics of small wavelength modes, which is not expected to play an important role on the large scale spoke mode of interest here. Further, the code is accelerated by MPI parallelization and the domain decomposition is used. Subcycling of electrons relative to ions is also used (11 times of electron sub-cycles per ion cycle) to reduce the numerical cost \cite{adam1982electron}. To the end, the steady state can be achieved at simulation time about $1.5 \mu s$, with the computation time about several hours for the cases under investigation.

 \begin{table*}[]
    \centering
\begin{tabular}{p{5.9cm}|p{3.6cm}}
% \hline
% \multicolumn{4}{|c|}{Table 1} \\
 \hline
 \hline
 Physical quantity & value  \\
 \hline
plasma density $n_{e0}$   & $\sim 1\times 10^{16}{\rm m^{-3}}$   \\
Electron temperature $T_e$   & $\sim 5{\rm eV}$  \\
Ion temperature $T_i$   & $\sim 1{\rm eV}$  \\
Electric field $E_0$ & $10^3-10^4 {\rm V/m}$ \\
Ion mass $m_i$ & $6.68 \times 10^{-26}{\rm kg}$ \\
Electron Debye length &  $\sim 0.1 {\rm mm}$ \\
Density gradient length $L_n$ & $\sim 0.5 {\rm mm}$ \\
Electron gyroradius $\rho_e$ & $\sim 0.15 {\rm mm}$ \\
Ion gyroradius $\rho_i$ & $\sim 1 {\rm cm}$ \\
Electron plasma frequency $\omega_{pe} $ & $1 \times 10^{10}{\rm rad/s}$\\
Electron collision rate $\nu_{en}$ & $\sim 2.5\times 10^8 s^{-1}$ \\
Electron thermal velocity $v_{e,th}$  & $\sim 10^6 {\rm m/s}$ \\
Ion radial velocity $v_{i0}$ & $1-10 \times 10^3 {\rm m/s}$ \\
Ion sound velocity $c_s$ & $\sim 3.5\times10^3 {\rm m/s}$ \\
${\bf E_{0}} \times {\bf B_0}$ drift $v_E$ & $5-50\times 10^{4} {\rm m/s}$ \\
Diamagnetic drift $v_d$ & $\sim 2.5\times 10^{5} {\rm m/s}$ \\
 \hline
 \hline
\end{tabular}
\hspace*{-1cm}\caption{The typical physical quantities and derived plasma parameters of interest during the nonlinear evolution of the spoke instability at and in the immediate neighborhood of the anode sheath in the PIC simulation with $B_0=40 {\rm mT}$}
\end{table*}

To identify the modes developed in the simulations, theoretical calculations in the frame of the gradient drift instability were made and compared with the simulated spectra. It is convenient to discuss the characteristics of the gradient drift instability with reference to the linear and local dispersion relation. In our model with homogeneous ${\bf B_0}$, the electrostatic waves are initiated in the anode sheath, where the electrons experience the diamagnetic drift and ${\bf E_0}\times {\bf B_0}$ drift, respectively:
\begin{equation}
\mathbf{v_d}=T_e\triangledown n_{e0} \times \mathbf{B_0}/eB_0^2n_{e0}=T_e/eB_0L_n {\bf y},
\end{equation}

\begin{equation}
\mathbf{v_E} = \mathbf{E_0} \times \mathbf{B_0}/B_0^2=E_0/B_0 {\bf y}.
\end{equation}

\noindent where $n_{e0}$ is the local equilibrium electron density, $e$ is the elementary charge and $L_n=n_{e0}/\triangledown n_{e0}$ is the electron density gradient length. Ions are unmagnetized and cold, and can be accelerated by $E_0$ to form the ion beam with mean velocity $v_{i0}$. With the consideration of quasi-neutrality, electron inertia, electron gyro-viscosity, Debye length effect and electron collisions, the two-dimensional two-fluid linear dispersion equation has the form \cite{smolyakov2016,Frias2012,Xu2021}:

\begin{equation}
(k_{y}^2+k_{x}^2)\lambda_{De}^2=\frac{(k^2_{y}+k^2_{x})c^2_s}{(\omega-k_xv_{i0})^2} - \frac{\omega_{d}+(k^2_{y}+k^2_{x})\rho^2_{e}(\omega-\omega_{E}+i\nu_{en})}{\omega-\omega_{E}+(k^2_{y}+k^2_{x})\rho^2_{e}(\omega-\omega_{E}+i\nu_{en})}.
\end{equation}

\noindent where $k_{x}$ and $k_y$ are the angular wave numbers, $\omega$ the angular frequency, $\lambda_{De}$ the Debye length, $\rho_{e}=(T_e/m_e)^{1/2}/\omega_{ce}$ the electron Larmor radius, $m_e$ the electron mass, $\omega_{ce}$ the electron cyclotron frequency, $c_s=(T_e/m_i)^{1/2}$ the ion sound speed, $m_i$ the ion mass, $\omega_{
d}=k_{y}v_{d}$, $\omega_E=k_{y}v_E$ and $\nu_{en}$ the electron-neutral collision frequency. Here, $k_{y}$ and $k_{x}$ are in the unit of ${\rm rad/m}$ and $\omega$ in ${\rm rad/s}$. The left hand side of Eq. 3 refers to the Debye length effect derived from Poisson equation, the first term on the right hand side is attributed to the ion inertial response and the second term on the right hand is related to the electron response in the fluctuation including the inertia effect and the gyro-viscosity effect (finite Larmor radius). However, in our simulations, the instability is found to take place in the anode sheath where quasi-neutrality is violated, and the radial wave number is much smaller than the azimuthal wave number, i.e., $k_x \ll k_y$. With the radial component neglected and the non-neutrality taken into account (see Appendix A for the derivation), Eq. 3 becomes

\begin{equation}
k_{y}^2\lambda_{De}^2=\frac{\alpha k^2_{y}c^2_s}{\omega^2} - \frac{\omega_{d}+k^2_{y}\rho^2_{e}(\omega-\omega_{E}+i\nu_{en})}{\omega-\omega_{E}+k^2_{y}\rho^2_{e}(\omega-\omega_{E}+i\nu_{en})}.
\end{equation}

\noindent where $\alpha=n_{i0}/n_{e0}$ represents the non-neutrality coefficient. Here, $n_{i0}$ is the local equilibrium ion density. Based on Eq. 4, Fig. 2 gives the theoretical predictions of the real part and the imaginary part (growth rate) of the frequency as a function of the wave number at different $E_0$, $L_n$ and $\alpha$ with typical values in our simulations and other quantities listed in Table 2 . 

\begin{figure}
\center
\hspace*{-1.5cm}\includegraphics[clip,width=1.2\linewidth]{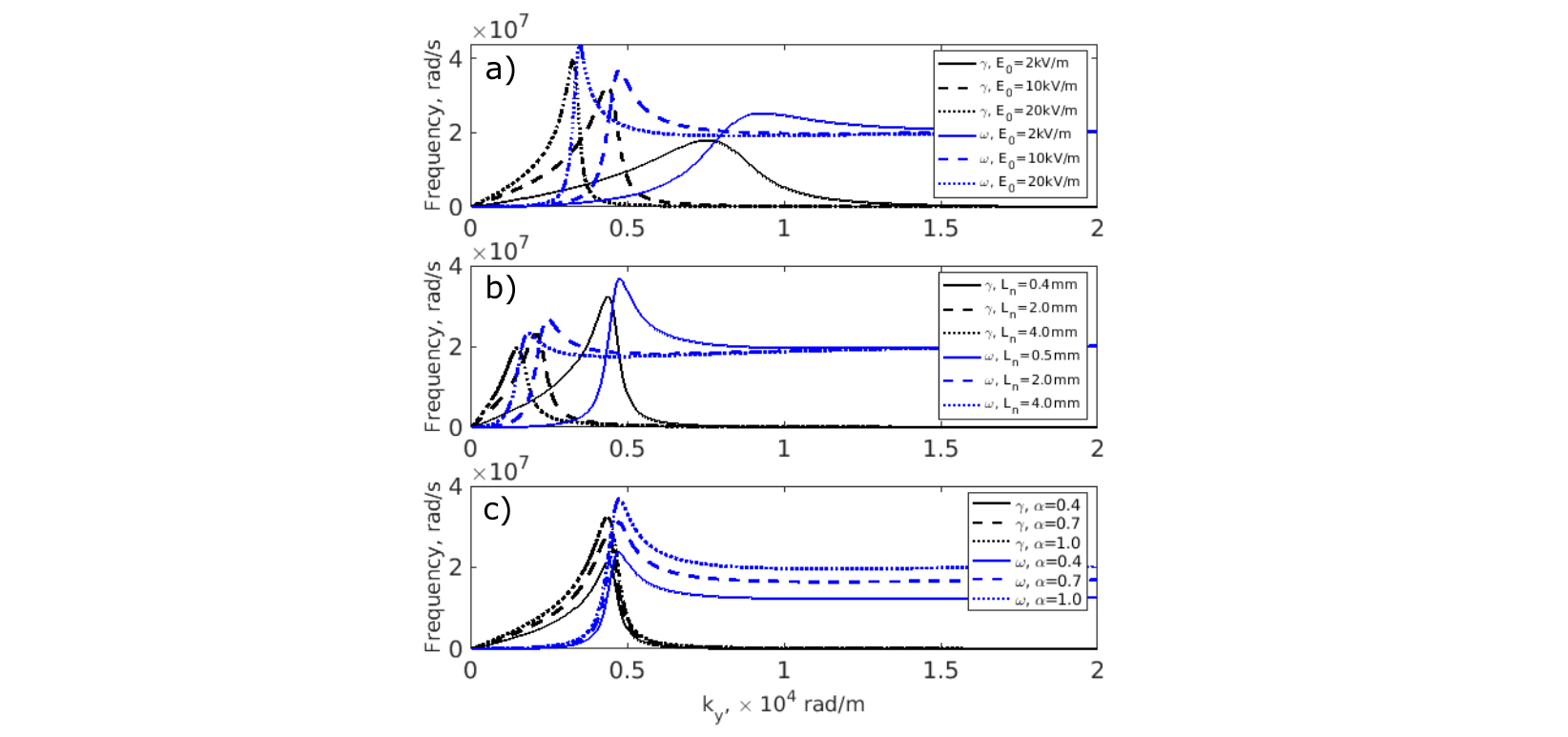}
\caption{The dispersion relation calculated from Eq. 4 when $B_0=40 {\rm mT}$, including the real frequency part and the imaginary part, a) for different values of $E_0$ with $L_n=0.4 {\rm mm}$, $n_{e0}=1\times 10^{16} {\rm m^{-3}}$ and $\alpha=1.0$, b) for different values of $L_n$ with $E_0=10 {\rm kV/m}$, $n_{e0}=1\times 10^{16} {\rm m^{-3}}$ and $\alpha=1.0$ and c) for different values of $\alpha$ with $E_0=10{\rm kV/m}$, $L_n=0.4 {\rm mm}$ and $n_{e0}=1\times 10^{16} {\rm m^{-3}}$.}
\end{figure} 

\begin{figure}
\center
\includegraphics[clip,width=0.5\linewidth]{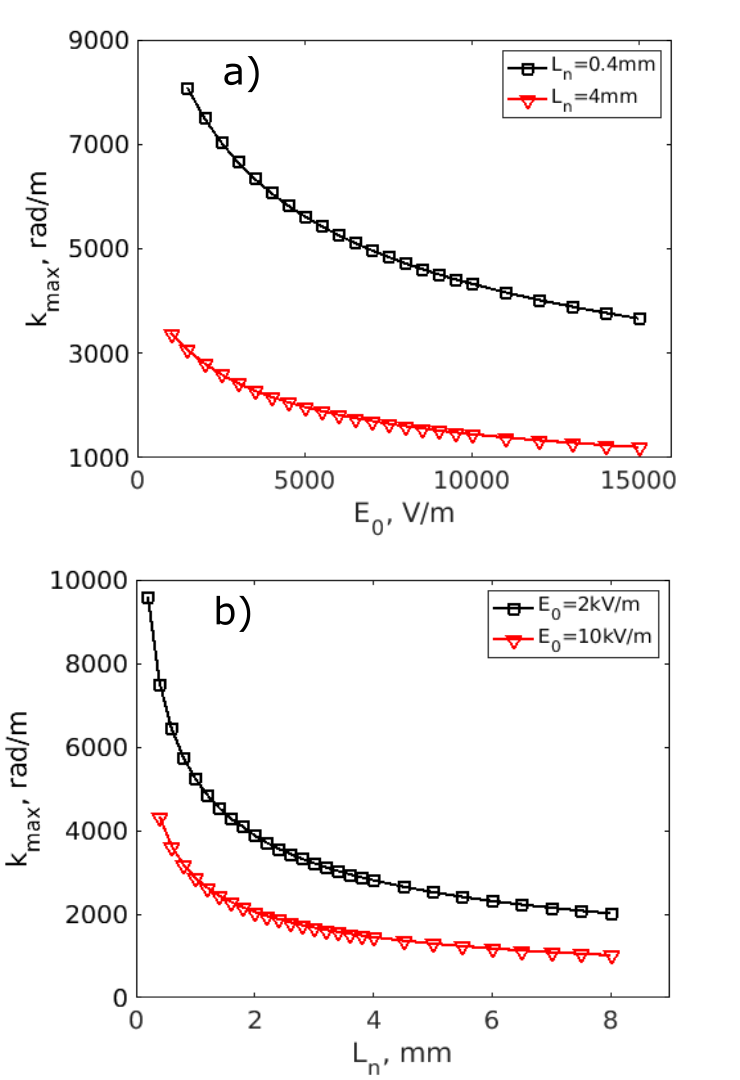}
\caption{According to Eq. 4, the calculated wave number $k_{max}$ of the mode having the largest grow rate as functions of a) $E_0$ and b) $L_n$. In the calculations, $n_{e0}=1\times 10^{16}m^{-3}$, $\alpha=1$ and $B_0=40 {\rm mT}$.}
\end{figure} 

According to the theory, different instability modes can develop. In the small wavelength regime $k_{y}\rho_{e}\gg 1$, the instability turns out to be the ion sound wave and Eq. 4 gives

\begin{equation}
\frac{\omega}{\omega_{pi}}=\frac{\alpha k_{y}\lambda_{De}}{\sqrt{1+k_{y}^2\lambda_{De}^2}}.
\end{equation}

If the quasi-neutrality is fulfilled ($\alpha=1$) and the Debye length effect can be further neglected, i.e., $k_{y}\lambda_{De}\ll 1$, Eq. 5 takes the simplest form $\omega=k_y c_s$. While in our cases with reduced plasma density, $k_{y}\lambda_{De}\gg 1$ may apply, particularly at the large k modes where Eq. 5 becomes $\omega=\omega_{pi}$. This scenario can be detected in Fig. 2a and 2b; The real frequency asymptotic to the ion plasma frequency $\omega_{pi}\approx 2 \times 10^7 {\rm rad/s}$ is observed when $k_y>1\times 10^4 {\rm rad/m}$. From Fig. 2, we note that growth rates of the ion sound instability (ISI) are negligible, attributed to the electron inertia effect and gyro-viscosity effect \cite{lakhin2018effects}.

In the long wavelength regime $k_{y}\rho_{e}\ll 1$, the Simon-Hoh instability (SHI) can be excited. With the Debye length effect neglected but non-neutrality accounted for, Eq. 4 reads:

\begin{equation}
\frac{\alpha k^2_{y}c^2_s}{\omega^2} = \frac{\omega_{d}}{\omega-\omega_{E}}.
\end{equation}

SHI is in nature the anti-drift wave destabilized by the the ${\bf E_0}\times {\bf B_0}$ drift and the well known criterion for SHI is the condition ${\bf E_0} \cdot \triangledown_x n_{e0}>0$ \cite{Simon1963,hoh1963}. 

When $(k_y\rho_e)^2\omega_E \simeq \omega_{d}$ and $k_y\rho_e\lesssim 1$, meaning the electron inertia starts to play a role and the gyro-viscosity effect can be negligible, SHI transits to lower hybrid instability (LHI) having the following dispersion relation:

\begin{equation}
k_{y}^2\lambda_{De}^2=\frac{\alpha k^2_{y}c^2_s}{\omega^2} - \frac{\omega_{d}+k^2_{y}\rho^2_{e}(\omega-\omega_{E}+i\nu_{en})}{\omega-\omega_{E}},
\end{equation}

From Eq. 7, destabilization sources of LHI can be ${\bf E_0} \times {\bf B_0}$ drift, diagmagnetic drift and electron collisions. If the destabilization sources and Debye length term are neglected, Eq. 7 comes to its simplest form $\omega^2=\alpha\omega_{ce}\omega_{ci}$ with non-neutrality effect considered.

It is noteworthy that the growth rates shown in Fig. 2 are all peaked at one single mode. For the nine cases shown in Fig. 2, $(k_{max}\rho_e)^2 k_{max}v_E/k_{max}v_{d}$ is in the range of $0.2-0.4$ and $k_{max}\rho_e$ is in the range of $0.2-1.2$, where $k_{max}$ is the wave number of the most unstable mode with the maximum growth rate. This proves that the electron inertia is important and gyro-viscosity effect can be neglected for the most unstable modes, which are hereby LHI. Another interesting point is that the variation of $\alpha$ can change the amplitudes of $\gamma$ and $\omega$, but most features of LHI and SHI are still retained. Particularly, the mode number $k_{max}$ of the most unstable mode of the primary interest here is unaltered with different non-neutrality coefficients as seen in Fig. 2c. From the calculation of Eq. 4, the wave number $k_{max}$ as functions of $E_0$ and $L_n$ is presented in Fig. 3a and Fig. 3b. From Fig. 3, one can see that with the increase of either $E_0$ or $L_n$, $k_{max}$ decreases, meaning the most unstable mode shifts from the short wavelength to the long wavelength. As seen later, this theoretical picture enables explain the mode transition towards the spoke formation observed in our simulations.

We should emphasize that for our model with uniform $B_0$ and $k_x \ll k_y$, the instability criterion ${\bf E_0}\cdot\triangledown_x n_{e0}>0$ applies not only in the long wavelength regime (SHI) but also in the small wavelength regime (LHI and ISI) \cite{lakhin2018effects}. 

\section{Results and discussions}

In the nonlinear saturated stage, our numerical simulations exhibit the rotating spoke phenomena as shown in Fig. 4, giving the x-y profiles of electron density, potential and electron temperature when $B_0=40 {\rm mT}$. It is clearly seen that the plasma is organized as an m=1 coherent structure and forms a potential hump region, which has higher potential compared to the surrounding region. As expected, electrons are heated in the double layer 
enclosing the potential hump region. In this case, the spoke rotates in the $+{\bf E}\times {\bf B}$ direction with the velocity approximately $7 {\rm km/s}$, in good agreement with cross field discharge experiments \cite{maass2021synchronising,Hecimovic_2016,Ito2015}. For the purpose of the present study, the nonlinear evolution of the micro fluctuations in the early phase of the simulations were studied to uncover the mechanism behind the spoke formation. Unless otherwise noted, in what follows, the magnetic field strength is $B_0=40 {\rm mT}$. The simulations for other magnetic fields in the range of $B_0=[20 {\rm mT}, 80{\rm mT}]$ were also conducted, and the basic physics shown below applies as well. In addition, $\langle X \rangle _y$ means the radially averaged quantity $X$ in the anode sheath x=[0, 1 {\rm mm}]; $\langle X \rangle _x$ means the azimuthally averaged quantity $X$ in the range of $y=[0, 20 {\rm mm}]$; $\langle X \rangle$ is the azimuthally and radially averaged quantity $X$ in the region $x=[0, 1 {\rm mm}]$ and $y=[0,20 {\rm mm}]$. We should point out that, in our simulations when $B_0>20 {\rm mT}$, the anode sheath is positive where ${\bf \langle E_x \rangle _x} \cdot \triangledown \langle n_{e}\rangle_x>0$ is satisfied, leading to the LHI excitement and the spoke formation. When $B_0<20 {\rm mT}$, the anode sheath is negative (${\bf \langle E_x \rangle _x} \cdot \triangledown \langle n_{e}\rangle_x<0$) and the simulations do not exhibit visible fluctuations and spokes are not observed.

\begin{figure}
\center
\includegraphics[clip,width=0.7\linewidth]{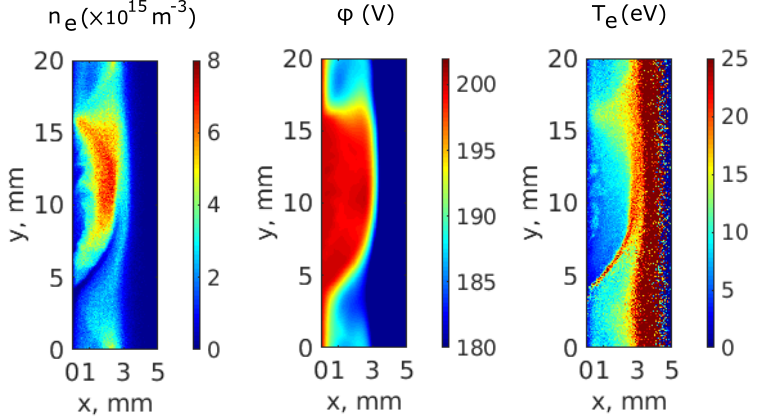}
\caption{Observation of the spoke in the nonlinear saturated stage. More pictures showing the saturated spoke can be found in our previous paper with the cylindrical model \cite{Xu2021}.}
\end{figure}

\begin{figure}
\center
\includegraphics[clip,width=1.0\linewidth]{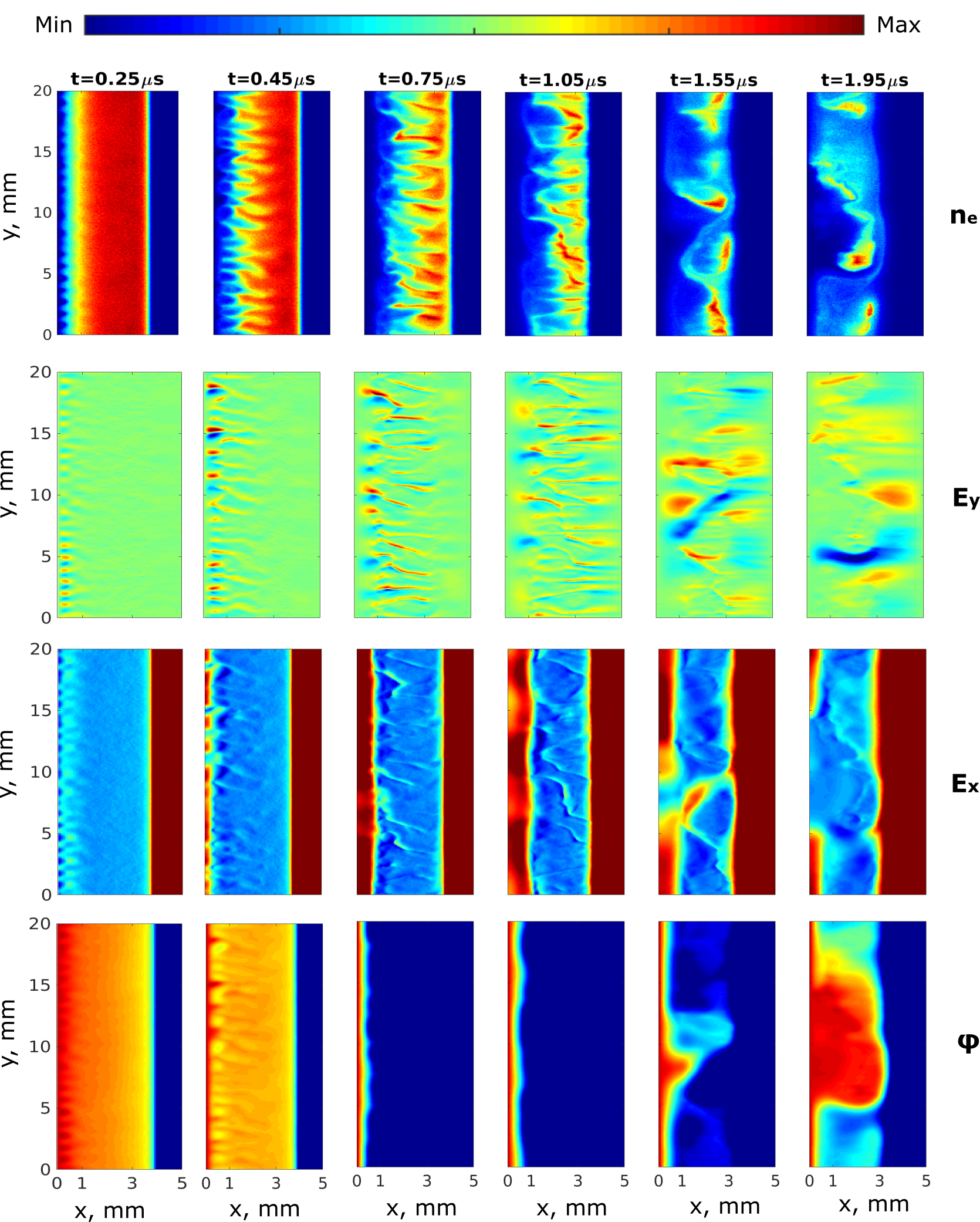}
\caption{Evolution of $n_e$, $E_{y}$, $E_x$ and $\varphi$ in the time period $[0.25{\rm \mu s}, 1.95{\rm \mu s}]$, during which the initial m=21 lower hybrid mode in the anode sheath transits to the m=1 rotating spoke mode. For $n_e$, the color bar range is $[0,4.1\times10^{16} {\rm m^{-3}}]$ at $t=0.25 {\rm \mu s}, 0.45 {\rm \mu s}, 0.75 {\rm \mu s}$, $[0,3.6\times10^{16} {\rm m^{-3}}]$ at $t=1.05 {\rm \mu s}$, $[0,2.4\times10^{16} {\rm m^{-3}}]$ at $t=1.55 {\rm \mu s}$ and $[0,1.45\times10^{16} {\rm m^{-3}}]$ at $t=1.95 {\rm \mu s}$. For $E_y$, the colorbar range is $[-1\times10^4 {\rm V/m}, 3\times10^4 {\rm V/m}]$. For $E_x$, the colorbar range is $[-1.5\times10^4 {\rm V/m}, 1.5\times10^4 {\rm V/m}]$. For $\varphi$, the colorbar range is $[180 {\rm V}, 202 {\rm V}]$.}
\end{figure}

\begin{figure}
\center
\includegraphics[clip,width=0.5\linewidth]{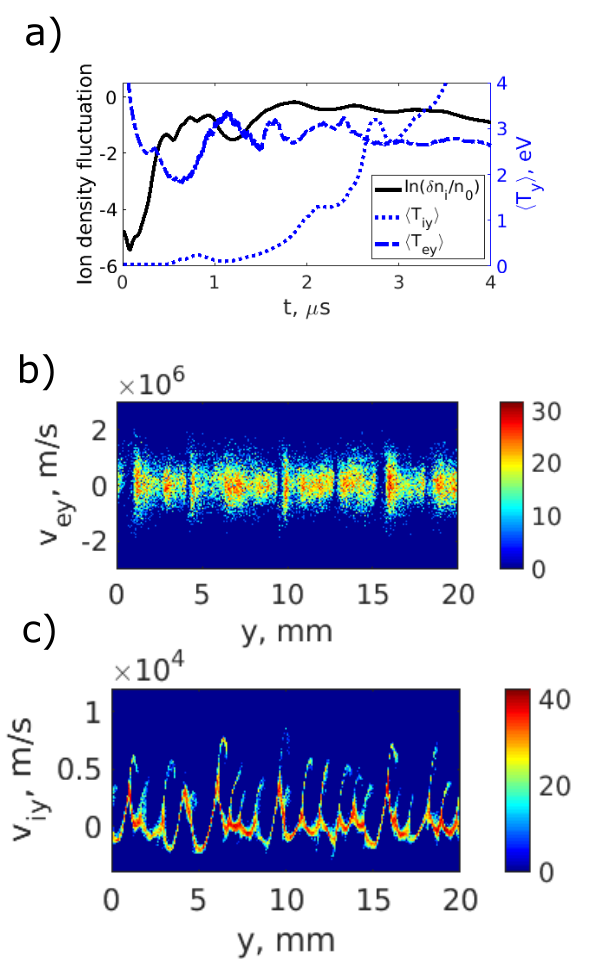}
\caption{a) Ion density fluctuation, together with azimuthal electron temperature $\langle T_{ey} \rangle$ and ion temperature $\langle T_{iy} \rangle$, as a function of time; b) and c) are electron and ion phase planes ($v_{ey}-y$ and $v_{iy}-y$) in the anode sheath $x=[0 {\rm mm}, 1 {\rm mm}]$ at the snapshot $t=0.7 {\rm \mu s}$. }
\end{figure}

\subsection{Onset and nonlinear evolution of LHI}
Chronologically, the evolution of the simulated electron density $n_e$, azimuthal electric field $E_{y}$, radial electric field $E_x$ and potential $\varphi$ goes as follows (see Figs. 5). 
Firstly, when the positive anode sheath starts to be built at the snapshot $t=0.25{\rm \mu s}$, the most unstable mode grows linearly inside the anode sheath and the fluctuation is seen in all parameters. After the instability saturation, the modes propagate towards the cathode (see $n_e$ and $E_y$ at snapshots $t=0.45{\rm \mu s}, 0.75{\rm \mu s}, 1.05{\rm \mu s}$). At $t=0.75{\rm \mu s}$, the positive anode sheath is almost fully established and with time increasing, the small scale mode filaments merge in the plasma bulk, leading to the formation of the large scale structures as seen in $n_e$ plots. At $t=1.55 {\rm \mu s}$ in the $E_x$ plot, a small section of the anode sheath collapses ($E_x\approx 0$ at $y\approx 8 {\rm mm}$). Subsequently, the collapse region expands azimuthally to the length of approximately $L_y/2$ (see $E_x$ at $t=1.95 {\rm \mu s}$).  Simultaneously, an equi-potential hump region with potential close to the anode potential forms as seen in $\varphi$ plot at $t=1.95 {\rm \mu s}$, when the double layer around the potential hump region is seen in $E_y$. 

\subsection{Saturation mechanism of LHI}

In order to have insight of the LHI saturation mechanism, the ion density fluctuation $\delta n_i/n_0$ in logarithm scale as a function of time is plotted in Fig. 6a, together with the azimuthal electron temperature $\langle T_{ey} \rangle$ and azimuthal ion temperature $\langle T_{iy} \rangle$. Here $\delta n_i=[\int^{x_2}_{x_1}\int^{y_2}_{y_1} n_i(x,y,t)^2 d x d y-(\int^{x_2}_{x_1}\int^{y_2}_{y_1}n_i(x,y,t) dx d y)^2]^{1/2}$ and $n_0=\langle n_i \rangle=\int^{x_2}_{x_1}\int^{y_2}_{y_1} n_i(x,y,t) d x d y$, where $y_1=0$, $y_2=L_y=20{\rm mm}
$, $x_1$ is the anode location $x_1=0 {\rm mm}$ and $x_2$ is the location of the anode sheath edge roughly $x_2=1.0 {\rm mm}$. From Fig. 6a, the instability growth saturates at about $t=0.4 {\rm \mu s}$ and the saturation is accompanied by the increase of $\langle T_{iy} \rangle$ from room temperature (initial condition) to several ${\rm eV}$ in the nonlinear stage. On the other hand, $\langle T_{ey} \rangle$ keeps almost unchanged at about $\langle T_{ey} \rangle  \approx 2.5{\rm eV}$ in the entire nonlinear stage. This clearly suggests that the instability saturation is due to the ion-wave trapping. To further verify this point, the electron and ion phase planes ($v_{ey}-y$ and $v_{iy}-y$) at $t=0.7{\rm \mu s}$ in the anode sheath are plotted in Fig. 6b and 6c respectively. One can see ions are trapped in the potential waves, while electron trapping is not observed.

\begin{figure}
\center
\includegraphics[clip,width=0.7\linewidth]{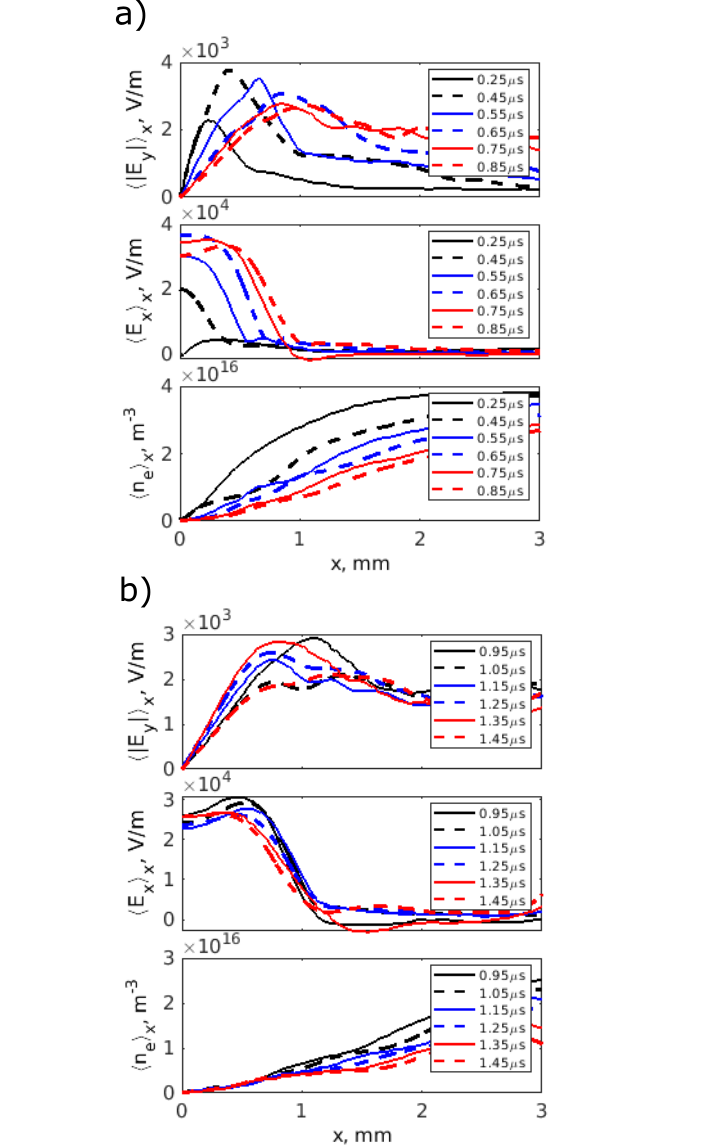}
\caption{Radial profiles of a) $\langle|E_{y}|\rangle_x$, b) $\langle E_x\rangle_x$ and c) $\langle n_e\rangle_x$ at different snapshots in the time period $t=[0.25 {\rm \mu s}, 1.45 {\rm \mu s}]$.}
\end{figure}

\begin{figure}
\center
\includegraphics[clip,width=1.0\linewidth]{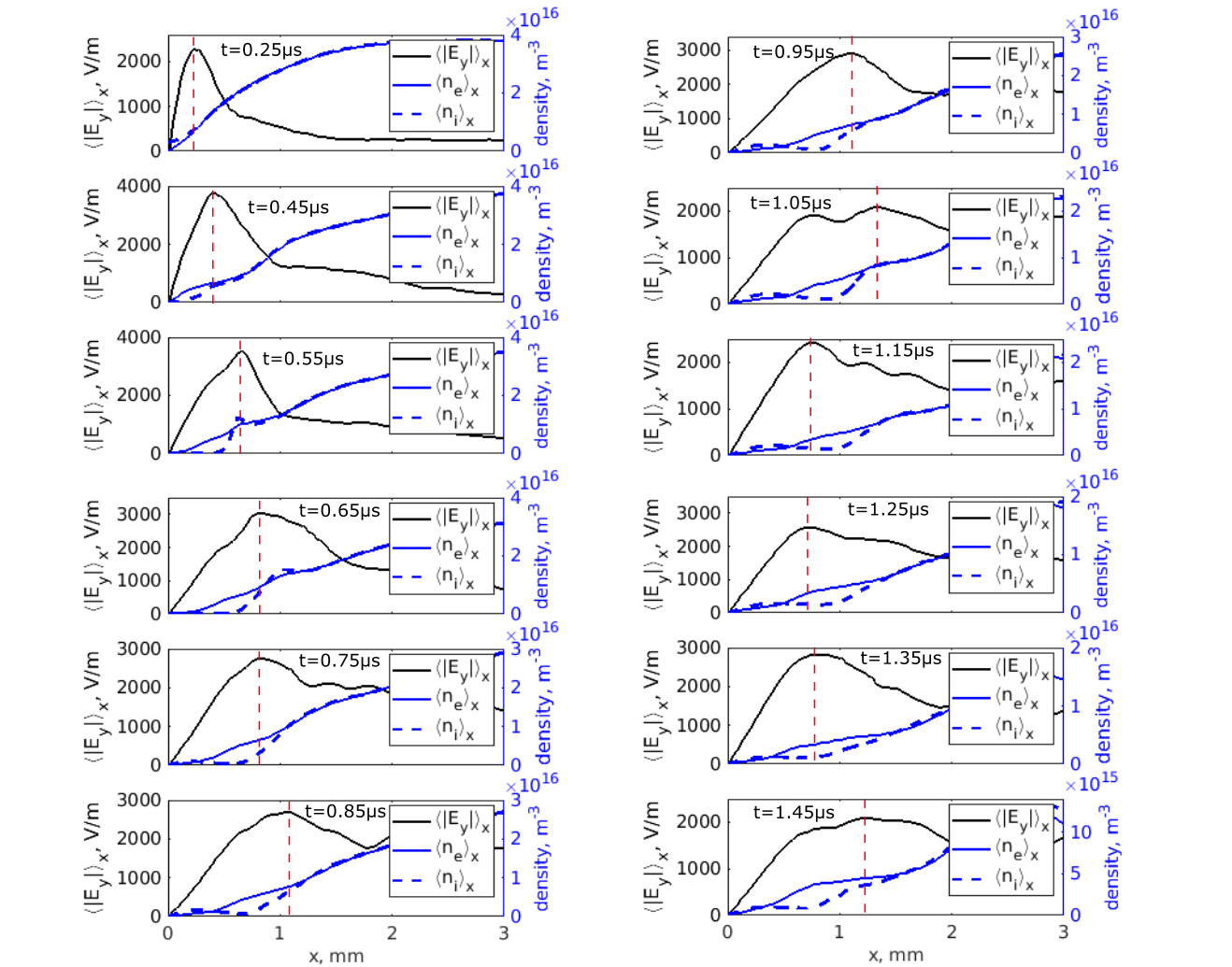}
\caption{Radial profiles of $\langle|E_{y}|\rangle_x$, $\langle n_e\rangle_x$ and and $\langle n_i\rangle_x$ in the near anode region at different snapshots. The red dashed lines represent the peak positions of $\langle|E_y|\rangle_x$, near which the quasi-neutrality is violated and the most unstable modes are initiated. }
\end{figure}

\begin{figure}
\center
\includegraphics[clip,width=0.9\linewidth]{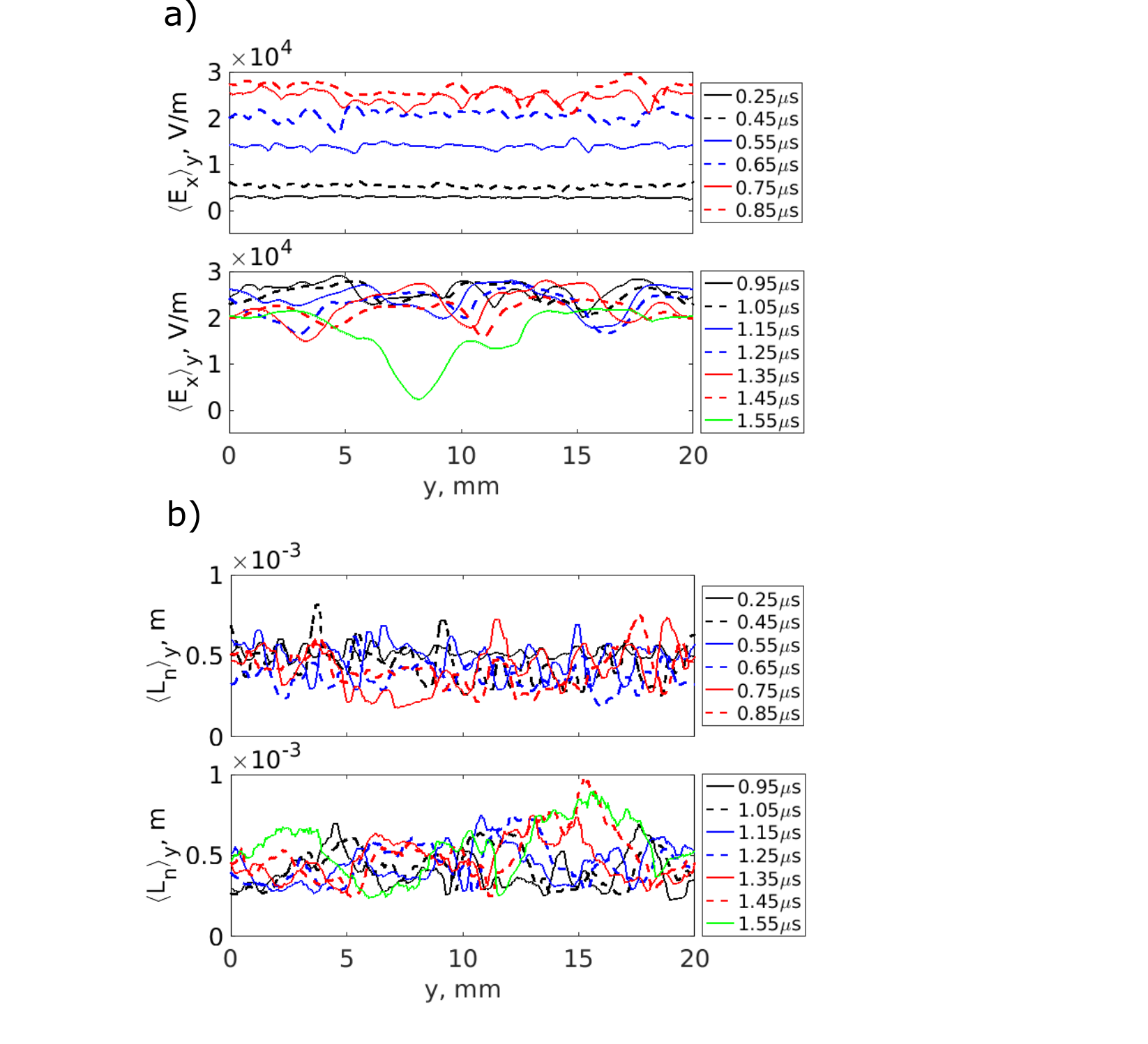}
\caption{Azimuthal profiles of a) $\langle E_{x}\rangle_y$ and b) $\langle L_n\rangle_y$ at different snapshots.}
\end{figure}

\begin{figure}
\center
\includegraphics[clip,width=0.6\linewidth]{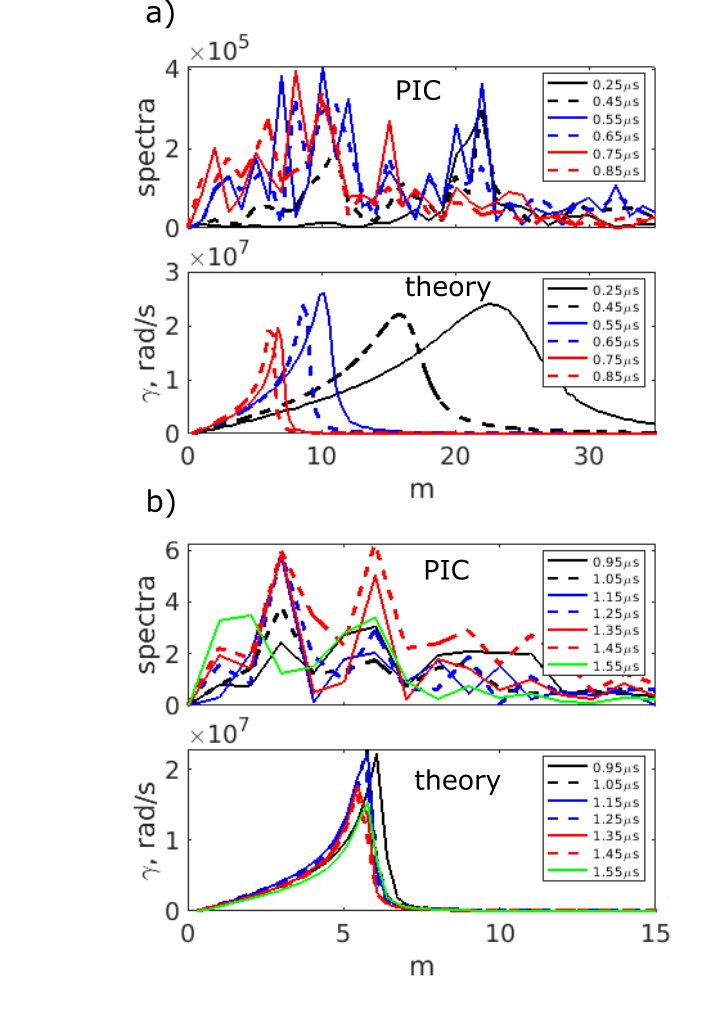}
\caption{The anode sheath $E_y$ spectra in the azimuthal direction using FFT, in comparison with the theoretical predictions of grow rates as a function of $m$ in the time period a) $t=[0.25 \mu s, 0.85 \mu s]$, and b) $t=[0.95 \mu s, 1.55 \mu s]$. }
\end{figure}

\begin{figure}
\center
\includegraphics[clip,width=0.6\linewidth]{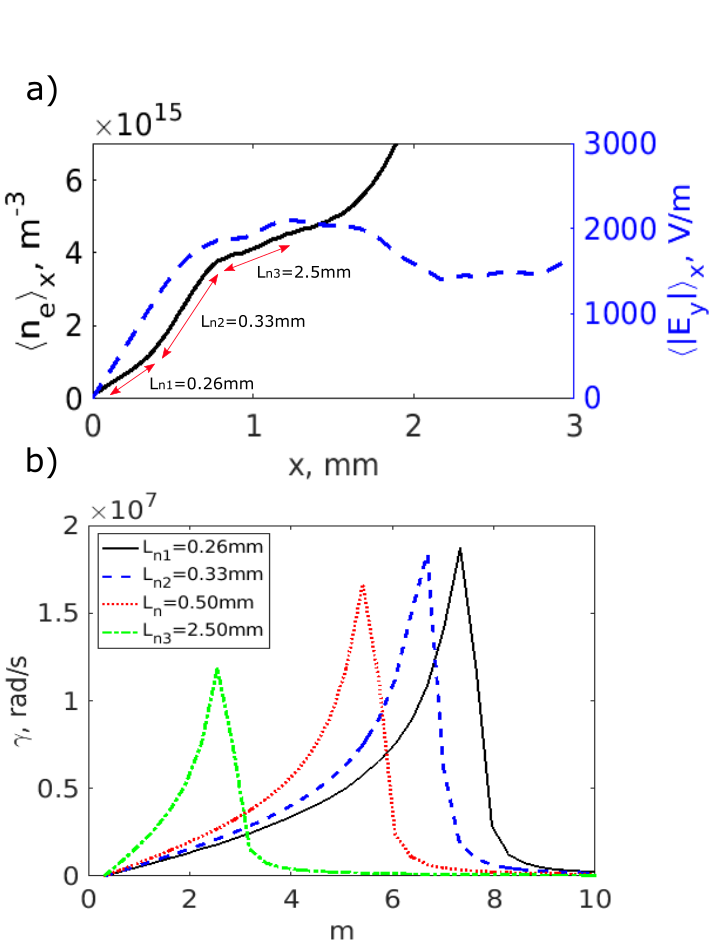}
\caption{a) Radial profiles of $\langle n_e\rangle_x$ and $\langle|E_{y}|\rangle_x$ in the near anode region at the snapshot $t=1.45 {\rm \mu s}$, and b) theoretical grow rates as a function of $m$ using different values of $L_n$ with $E_0=21.5 {\rm kV/m}$ and $n_{e0}=2.0\times10^{15} {\rm m^{-3}}$ listed in Table 3 at $t=1.45 {\rm \mu s}$.}
\end{figure}

 \begin{table*}[]
    \centering
\begin{tabular}{p{1.0cm}|p{1.5cm}|p{0.8cm}|p{1.8cm}|p{0.8cm}}
 \hline
 \hline
$t({\rm \mu s})$ & $E_0(\times 10^3)$ & $L_n$ & $n_{e0} (\times 10^{15})$ & $\alpha$\\
 \hline
$0.25$  & 3.0  & 0.52 & 14.7 & 1.02 \\
$0.45$  & 5.4 & 0.45  &  7.8 & 0.84\\
$0.55$  & 14.0 & 0.48 &  6.4 & 0.73\\
$0.65$  & 20.6 & 0.37 &  4.7 & 0.52\\
$0.75$  & 24.7 & 0.42 &  3.4 & 0.42\\
$0.85$  & 26.0 & 0.42 &  2.8 & 0.49\\
$0.95$  & 25.7 & 0.40 &  2.4 & 0.58 \\
$1.05$  & 25.1 & 0.43 &  2.2 & 0.69\\
$1.15$  & 24.1 & 0.44 &  2.1 & 0.74\\
$1.25$  & 22.8 & 0.45 &  2.0 & 0.64\\
$1.35$  & 22.7 & 0.47 &  2.0 & 0.54\\
$1.45$  & 21.5 & 0.50 &  2.0 & 0.47\\
$1.55$  & 17.1 & 0.55 &  1.9 &0.43\\
 \hline
 \hline
\end{tabular}
\hspace*{-1cm}\caption{The table gives the values of parameters $E_0$, $L_n$, $n_{e0}$ and $\alpha$ derived from the PIC/MCC simulations for calculating the instability dispersion relation using Eq. 4. The units of $E_0$,$L_n$ and $n_{e0}$ are ${\rm V/m}$, ${\rm mm}$ and ${\rm m^{-3}}$}
\end{table*}

\subsection{Mode transition in the nonlinear evolution}

As described in Fig. 5, the spoke formation is accompanied by the nonlinear transition from an initial small wavelength modes to long wavelength modes in the period $t=[0.25 {\rm \mu s},1.95 {\rm \mu s}]$. Fig. 10 presents the $\langle E_y\rangle_y$ spectra in the azimuthal dimension using fast Fourier transform at different snapshots. As seen in Fig. 10a and 10b, in the linear stage at $t=0.25 {\rm \mu s}$, the $m=21$ mode is dominant. With time increasing, the small scale mode undergoes a sequence of transitions to longer wavelength modes, and the shorter wavelength modes and the longer wavelength modes can coexist during the transition phase. It is noted that the mode transition takes place in the early phase of the simulation, when the steady state is not achieved. Therefore, the mode transition may be related to the evolution of the electric field and electron density during the establishment of the positive anode sheath. Shown in Fig. 7 are the radial profiles of $\langle|E_y|\rangle_x$, $\langle E_x\rangle_x$ and $\langle n_e\rangle_x$ at different snapshots. As seen in Fig. 7, with time increasing from $0.25 {\rm \mu s}$ to $0.75 {\rm \mu s}$, $n_e$ decreases and $E_x$ increases and afterwards they keep almost unchanged, indicating the anode sheath is fully established at around $0.75 {\rm \mu s}$. It is also seen that the $\langle |E_y| \rangle _{x}$ peaks are in the anode sheath region. In order to estimate the non-neutrality effect in the vicinity of the instability position, Fig. 8 presents the detailed radial profiles of $\langle n_e \rangle _{x}$, $\langle n_i \rangle _{x}$ and $\langle |E_y| \rangle _{x}$ at each snapshot in the time period $t=[0.25 {\rm \mu s}, 1.45{\rm \mu s}]$. The red dashed lines denote the locations of the $\langle |E_y| \rangle _{x}$ peaks in Fig. 8, corresponding to the approximate locations of the unstable modes. Fig. 7 and Fig. 8 clearly provide the evidence that the instability takes place in the anode sheath, where the quasi-neutrality is violated and the condition ${\bf \langle E_x \rangle _x} \cdot \triangledown \langle n_{e}\rangle_x>0$ is satisfied. 

We attempt to use the linear and local theory (Eq. 4) to explain the mode transition in the nonlinear stage. There are several crucial local parameters in the theory, $E_0$, $L_n$, $n_{e0}$ and $n_{i0}$ which are directly derived from the simulations. In our calculations, $E_0=\langle E_x \rangle$, $n_{e0} = \langle n_e \rangle$ and $n_{i0}=\langle n_i \rangle$. $L_n$ is obtained by three steps: 1) radially averaged $n_e$ and $\triangledown_x n_e$ gives $\langle {n_e} \rangle _{y}$ and $\langle {\triangledown_x n_{e}} \rangle _{y}$; 2) $\langle {n_e}\rangle _y/\langle{\triangledown_x n_{e}}\rangle _y$ gives the azimuthally dependent $\langle L_{n}\rangle _y$; 3) azimuthally averaged $\langle L_{n}\rangle _y$ approximates $L_n$. With $n_{e0}$ and $n_{i0}$, $\alpha=n_{i0}/n_{e0}$ is thus derived. $E_0$, $L_n$, $n_{e0}$ and $\alpha$ are listed in Table 3 for different snapshots. To evaluate the validity of the linear and local theory using $E_0$ and $L_n$ for the nonlinear analysis, we plotted the azimuthal profiles of $\langle E_x \rangle _y$ and $\langle L_{n} \rangle _y$ in Fig. 9. It is seen, at the most snapshots, the uncertainty of $\langle E_x \rangle _y$ and $\langle L_{n} \rangle _y$ are within $20\%$, hence the linear theory is expected to be applicable and render quantitative insights. Inserting these values of Table 3 in Eq. 4, the theoretical growth rates as a function of mode number $m$ are presented in Fig. 10a and 10b, in comparison with the simulated spectra. It is seen that the most unstable modes predicted by the theory are in good agreements with the most pronounced modes in simulated spectra from $t=0.25 {\rm \mu s}$ to $t=0.75{\rm \mu s}$ in Fig. 10a. However, in Fig. 10b after $t=0.75 {\rm \mu s}$, the theory predicts that the dominant mode should be stably $m \approx 6$, not consistent with the simulations where the mode further jumps to $m=3,2,1$. The most likely reason for the disagreement is that $E_x$, $L_n$ and $n_e$, are radially dependent, resulting in the local theory inaccurate in predicting the dominant mode. For example, shown in Fig. 11a are the radial profiles of $\langle n_e\rangle_x$ and $\langle|E_{y}|\rangle_x$ at the snapshot $t=1.45 {\rm \mu s}$. We see in the anode sheath, the electron density exhibits three branches giving different density gradient length, $L_{n1}=0.26 {\rm mm}$, $L_{n2}=0.33 {\rm mm}$ and $L_{n3}=2.5 {\rm mm}$ (the averaged value $L_n=0.5 {\rm mm}$ is shown in Table 3). Using these values with $E_0$ and $n_0$ in Table. 3, the theoretical growth rates as a function of $m$ are plotted in Fig. 11b. One can see that the dominant mode shifts from $m\approx6$ with $L_n=0.5 {\rm mm}$ to $m\approx2$ with $L_{n3}=2.5 {\rm mm}$, which is in a good agreement with the simulations in Fig. 10b. Besides, from the peak of $\langle|E_y|\rangle_x$ in Fig. 11a, the instability indeed occurs at the radial location where $L_{n3}=2.5 {\rm mm}$. The other reasons for the disagreement between theory and simulations can be the non-linearity and non-locality in the azimuthal direction or the ion beam formation in the radial direction \cite{xu2020self}.

\subsection{The formation of spoke potential hump}

\begin{figure}
\center
\includegraphics[clip,width=1.0\linewidth]{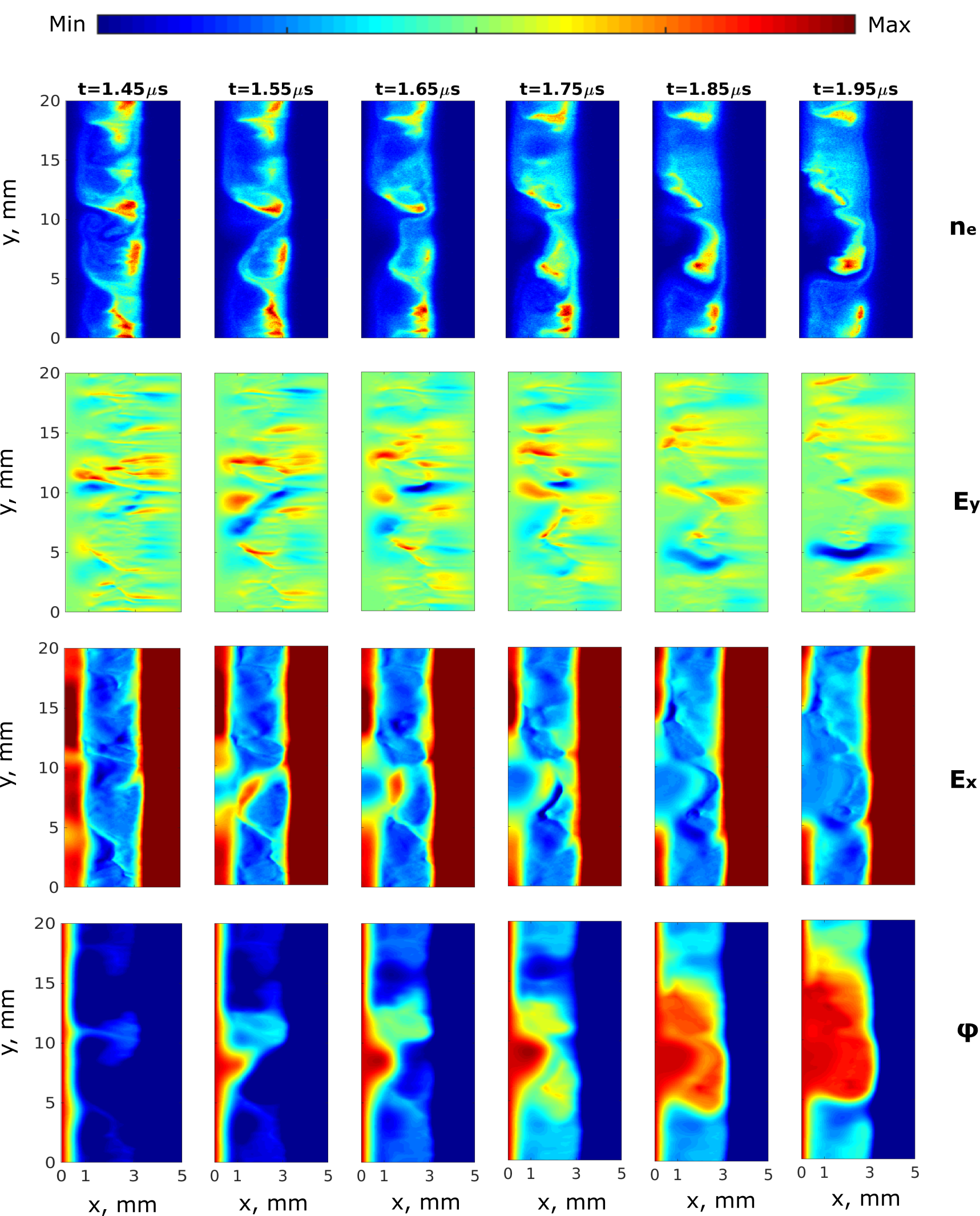}
\caption{Evolution of $n_e$, $E_{y}$, $E_x$ and $\phi$ in the time period $[1.45{\rm \mu s}, 1.95{\rm \mu s}]$, during which the images show how the spoke potential hump region is formed. For $n_e$, the colorbar range is $[0,2.3\times10^{16} {\rm m^{-3}}]$ at $t=1.45 {\rm \mu s}, 1.55 {\rm \mu s}, 1.65 {\rm \mu s}$, $[0,1.7\times10^{16} {\rm m^{-3}}]$ at $t=1.75 {\rm \mu s}, 1.85 {\rm \mu s}$ and $[0,1.45\times10^{16} {\rm m^{-3}}]$ at $t=1.95 {\rm \mu s}$. For $E_y$, the colorbar range is $[-1\times10^4 {\rm V/m}, 3\times10^4 {\rm V/m}]$. For $E_x$, the colorbar range is $[-1.5\times10^4 {\rm V/m}, 1.5\times10^4 {\rm V/m}]$. For $\varphi$, the colorbar range is $[180 {\rm V}, 202 {\rm V}]$.}
\end{figure}

\begin{figure}
\center
\includegraphics[clip,width=0.7\linewidth]{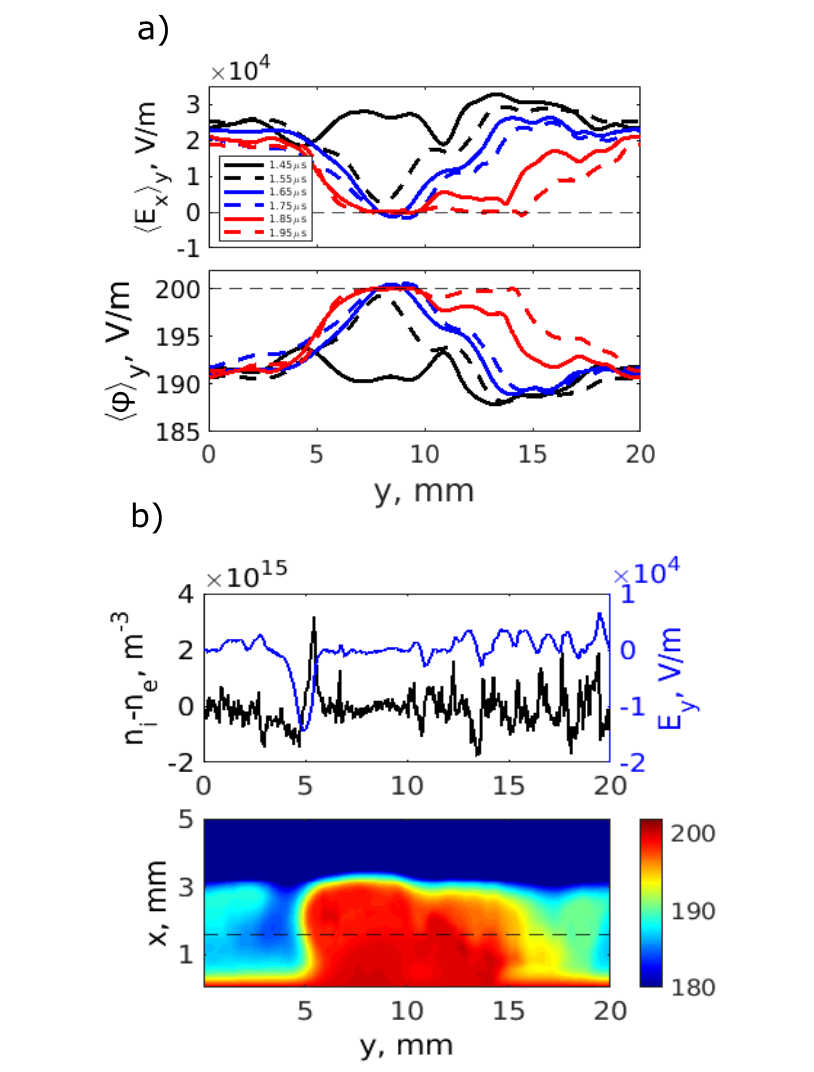}
\caption{a) azimuthal profiles of $\langle E_{x}\rangle_y$ and $\langle \varphi \rangle_y$ in the anode sheath at different snapshots displays how the potential hump is formed. b) azimuthal profiles of space charge density ($n_i-n_e$) and $E_y$ are shown at the snapshot $t=1.95 {\rm \mu s}$ and at the location $x=1.6 {\rm mm}$ denoted by the dashed line in the $\varphi$ figure below. The pronounced double layer is seen in front of the potential hump region.}
\end{figure}

The mode transition addressed above using linear theory approximation does not explain the presence of the highly nonlinear spoke mode. Fortunately, the simulation allows us to visualize the spoke formation to gain physical insights. To specify, the evolution of $n_e$, $E_{y}$, $E_{x}$ and $\varphi$ in the duration $t=[1.45 {\rm \mu s}, 1.95 {\rm \mu s}]$ is plotted in Fig. 12. As seen in the $\varphi$ and $E_x$ plots at $t=1.45 {\rm \mu s}$, the positive anode sheath (with positive $E_x$) fully covers the anode. At $t=1.55 {\rm \mu s}$, the $m=1$ mode is induced and the growth of the mode fluctuation results in the trough of $E_x$ becoming approximately zero at $y \approx 8 {\rm mm}$, implying the positive anode sheath collapses. As the growth continues, the anode sheath collapse region expands azimuthally and simultaneously the $m=1$ mode propagates radially (see $\varphi$ plots at $t=[1.55 {\rm \mu s},1.95 {\rm \mu s}]$). To the end, a potential hump region occupying the half azimuthal dimension with quasi-equipotential close to the anode potential is formed and seen in the $\varphi$ and $E_x$ plots at $t=1.95{\rm \mu s}$. As a result, the boundary of the potential hump region exhibits a potential drop linking the lower potential region. Fig. 13a further gives the azimuthal profiles of the $\langle E_x\rangle_y$ and $\langle \varphi \rangle_y$ in the anode sheath. It is clearly seen that $\langle E_x \rangle_y$ trough of the $m=1$ mode declines with time until getting zero and then expands azimuthally leading to the collapse of the half anode sheath. Correspondingly, the potential peak approaches the anode potential and then azimuthally expands to form the large scale potential hump region; The double layer can be discerned at the edge of the potential hump region in Fig. 13b. With the large electric field generated in the double layer, electrons are therefore energized due to collisional heating, resulting in the locally enhanced ionization \cite{Xu2021,Boeuf2014}. One interesting thing from Fig. 12 and Fig. 13 is that, when the spoke is established, the electric field generated in the double layer $E_{DL}$ is approximately equal to the radial electric field $\langle E_x \rangle$ in the anode sheath (prior to its collapse), $\langle E_x \rangle \approx E_{DL}\approx 1.5 \times 10^4 {\rm V/m}$ in this particular case. In fact, this is also true at different magnetic fields. It is therefore possible to obtain $E_{DL}$ from $\langle E_x \rangle$, which can be calculated in the framework of the classic/collisional theory. With $E_{DL}$ known, we can further quantify the spoke associated electron transport and heating, which is reserved for our future study.

\subsection{Model with negative anode sheath}
We have shown that the mode transition and the formation of the potential hump region are closely related with the dynamics of the positive anode sheath. Therefore, it is interesting to see the scenario when the anode sheath is negative, i.e., the plasma potential is larger than the anode potential. To do so, we modified the model by removing the magnetic field at the near anode region:

\begin{equation}
B_0=0 {\rm mT}   (0 {\rm mm}<x<1 {\rm mm});
\\
B_0=40 {\rm mT}  (1 {\rm mm}<x<5 {\rm mm})
\end{equation}

\begin{figure}
\center
\includegraphics[clip,width=0.9\linewidth]{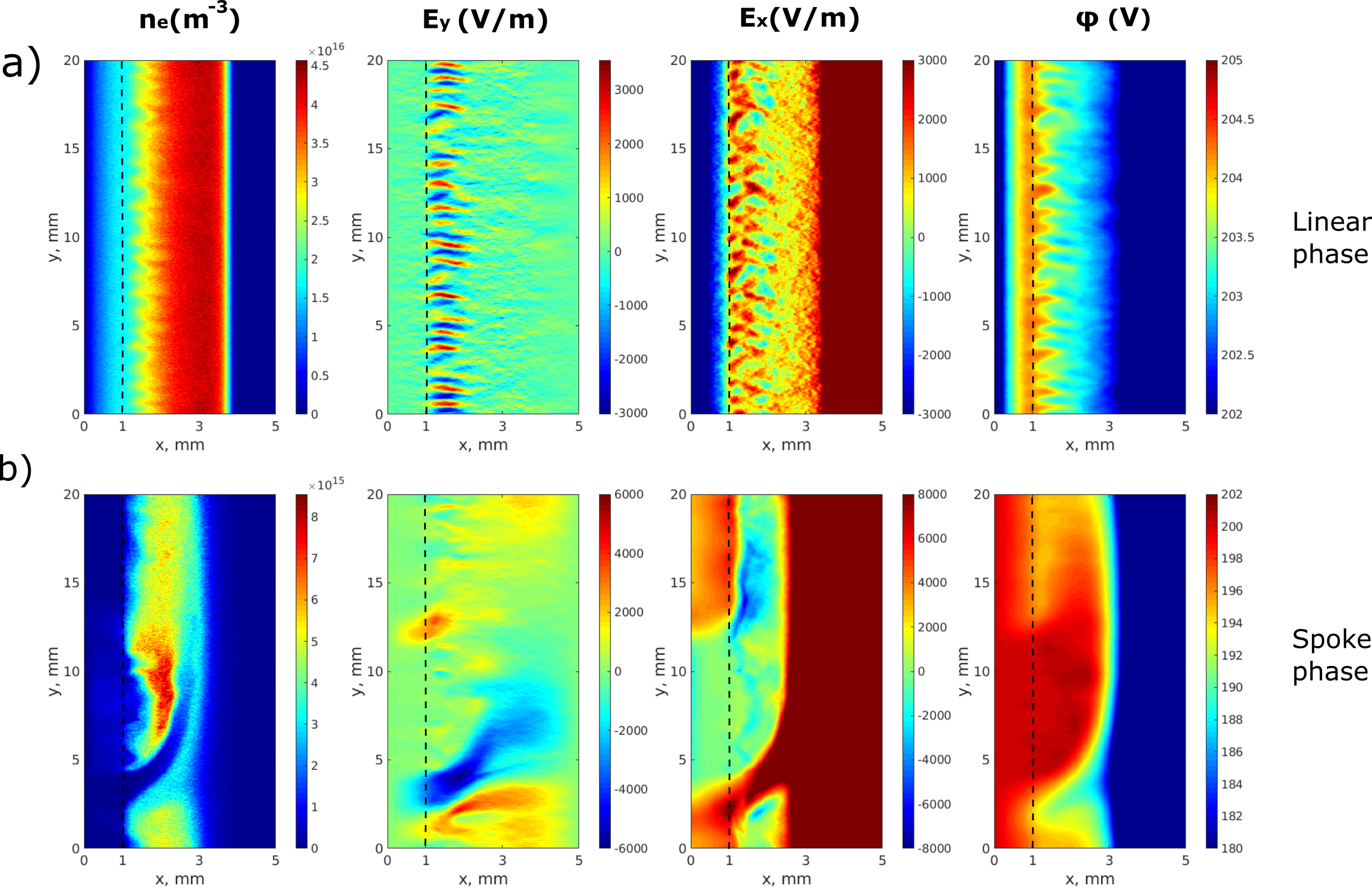}
\caption{Radial-azimuthal profiles of $n_e$, $E_{y}$, $E_x$ and $\varphi$ a) in the linear stage and b) in the saturated spoke stage.}
\end{figure}

Running the modified model outputs the x-y profiles of $n_e$, $E_{y}$, $E_{x}$ and $\varphi$ in the linear stage and the saturated spoke stage shown in Fig. 14a and 14b, respectively. In the linear stage, from $E_x$ and $n_e$ plots, it is seen that $E_x$ is negative in the anode sheath (thus ${\bf \langle E_x \rangle _x} \cdot \triangledown \langle n_{e}\rangle_x<0$), and the instability is not excited there (see $E_y$ plot). Instead, the instability is initiated in front of the $B$ cutoff position, $1{\rm mm}<x<2{\rm mm}$, where the local $E_x$ is positive and thus ${\bf \langle E_x \rangle _x} \cdot \triangledown \langle n_{e}\rangle_x>0$. From the $\varphi$ plot, at the $B$ cutoff position ($x=1 {\rm mm}$), there exists a potential peak we here refer to as a 'virtual anode'. In the saturated spoke stage shown in Fig. 13b, the m=1 spoke mode is clearly seen, along with the virtual anode sheath collapse (see the plot of $E_x$ at $2.5 {\rm mm}<y<12.5 {\rm mm}$). Due to the virtual anode sheath collapse, the spoke potential hump region with the surrounding double layer is formed (see the $\varphi$ and $E_y$ plots). Again, it is shown that the electric field in the double layer $E_{DL}$ approximately equals to the radial electric field $\langle E_x \rangle _y$ in the virtual anode sheath (before its collapse) $E_{DL} \approx \langle E_x \rangle _y \approx 4000 {\rm V/m}$. In this case, the nonlinear transition from short wavelength modes to long wavelength modes in the virtual anode sheath ($1{\rm mm}<x<2{\rm mm}$) also occurs (not shown). Comparing the modified and unmodified models, we proved that the presence of the radial electric field either in the anode sheath or in the plasma bulk (virtual anode sheath), with the condition ${\bf \langle E_x \rangle _x} \cdot \triangledown \langle n_{e}\rangle_x>0$, suffices to destabilize the system and drive the formation of spoke. 

\section{Conclusion}

We studied the nonlinear dynamics and formation of an m=1 azimuthally rotating spoke mode under conditions of magnetically enhanced hollow cathode arc discharges using 2D radial-azimuthal fully kinetic PIC/MCC approach. The numerical model presumes a uniform plasma background initially, and the discharge is sustained by cathode emission electrons. In the steady state, our simulations exhibit robust spokes with the rotation velocity, potential hump and double layer consistent with experimental observations in typical cross field discharges, e.g., magnetrons and Hall thrusters. The onset and nonlinear evolution of the spoke instability can be backtraced and interpreted to elucidate the underlying mechanism driving the spoke formation.

For the present model with a uniform magnetic field, in both the early linear phase and nonlinear evolution phase, we showed that the instabilities occur in the positive anode sheath with the condition ${\bf \langle E_x \rangle _x} \cdot \triangledown \langle n_{e}\rangle_x>0$ fulfilled. To identify the instability modes developed in the simulations, the simulated spectra is compared with the two-fluid linear theory of the gradient drift instability modified with non-neutrality effects.  It is found that most features of the gradient drift instability are retained with non-neutrality effects. We show that the modes developed in simulations are in good agreement with the the most unstable modes predicted by the modified theory in both the linear phase and nonlinear evolution phase. Therefore, the modes developed in our simulations are identified as the lower hybrid type gradient drift instability (LHI).

%For the present model with a uniform magnetic field, in both the early linear phase and nonlinear evolution phase, we showed that the instabilities occurs in the positive anode sheath with the condition ${\bf \langle E_x \rangle _x} \cdot \triangledown \langle n_{e}\rangle_x>0$ fulfilled. To identify the instability modes developed in the simulations, the simulated spectra is compared with the two-fluid linear theory of the gradient drift instability. In our cases, we find the linear gradient drift instability theory is still applicable in predicting the instability dispersion relation in the nonlinear phase. The violation of quasi-neutrality in the anode sheath is accounted for in the theory by introducing a non-neutrality coefficient. It is found that most features of the gradient drift instability are retained with non-neutrality effects. We show that the modes developed in simulations are in good agreements with the the most unstable modes predicted by the modified theory in both the linear phase and nonlinear evolution phase. Therefore, the modes developed in our simulations are identified as the lower hybrid type gradient drift instability (LHI). 

The model reveals that the formation of the spoke potential hump region and the surrounding double layer are triggered by the local anode sheath collapse resulting from the LHI evolving into the long wavelength regime. The insight can be detailed as follows:
\par
1). At the early linear phase, when the radial electric field in the anode sheath becomes positive, the $m=21$ LHI mode is thereby initiated. The instability is associated with the fluctuations of plasma density, azimuthal electric field, radial electric field and potential. Subsequently, the instability can penetrate the plasma bulk and propagate towards the cathode.
\par
2). After the saturation of the linear growth, the nonlinear transition from short wavelength modes to long wavelength modes occurs in the anode sheath due to the evolution of the radial electric field and electron density gradient length during the establishment of the positive anode sheath. 
%The growth of the $E_r$ fluctuation results in zero or slightly negative $E_r$ in the trough of the m=1 mode, i.e., the local anode sheath collapse. Meanwhile, the mode propagates from the anode to the cathode due to the radial ion beam accelerated in the anode sheath. Further, the anode sheath collapse region expends azimuthally and radially, causing the formation of the potential hump region close to the anode potential covering the half azimuthal dimension and from anode to cathode presheath.
\par
3). With the full establishment of the positive anode sheath, the m=1 mode is thereby excited. The growth of the mode fluctuation results in the trough of the radial electric field becoming zero and the resulting anode sheath collapse region expanding azimuthally. As a consequence, the potential hump region covering the half azimuthal dimension with quasi-equipotential close to the anode potential is formed. The distortion of the electric potential surrounding the potential hump region leads to the formation of the double layer.

%The results also suggest that it is possible to derive the double layer electric field through the anode sheath (prior to its collapse) electric field. Therefore, our findings highlight the importance of the classical/collisional theory in predicting the electric field generated in the double layer and the spoke associated electron heating and transport, which is reserved for our future publications.

\section*{Acknowledgement}
The author A. Smolyakov was supported in part by NSERC Canada and the Air Force Office of Scientific Research FA9550-15-1-0226. Other authors have been supported by the German Science Foundation (DFG) within the SFB-TR 87 project framework and by the Research Department ‘Plasmas with
Complex Interactions’ of Ruhr University Bochum.

\appendix
\section{Derivation of the gradient drift instability dispersion relation with non-neutrality considered}

The following equations are calculated with Cartesian coordinate and only the mode developed in the azimuthal direction ($\mathbf{y}$) is accounted for as stated in the text. Ions are unmagnetized. The governing equations for cold ions are the mass and momentum conservation equations:

\begin{equation}
     \frac{\partial n_i}{\partial t} +  \triangledown \cdot (n_i \mathbf{v_i}) = 0,
\end{equation}

\begin{equation}
     \frac{\partial \mathbf{v_i}}{\partial t} +  (\mathbf{v_i} \cdot \triangledown)  \mathbf{v_i} = -\frac{e}{m_i}\triangledown \varphi.
\end{equation}

For linear perturbations, $\widetilde{n_i},\varphi,\widetilde{v_i} \sim exp[-i(\omega t-k_{y}y)]$, the system of Eq. A1 and Eq. A2 reduces to 

\begin{equation}
     \frac{\partial \widetilde{n_i}}{\partial t} + n_{i0} \frac{\partial \widetilde{v_i}}{\partial y}  = 0,
\end{equation}

\begin{equation}
     \frac{\partial \widetilde{v_i}}{\partial t}  = -\frac{e}{m_i}    \frac{\partial \varphi}{\partial y}.  
\end{equation}

\noindent where $n_{i0}$ is the equilibrium ion density. Note that in the local approximation, the equilibrium profiles can be considered constant ($\partial_y n_{i0}=\partial_y v_0=0$ where $v_0$ is the equilibrium ion velocity). From A3 and A4, the ion density perturbation $\widetilde{n_i}$ in response to the potential fluctuation $\varphi$ can be derived:

\begin{equation}
\frac{\widetilde{n_i}}{n_{i0}}=\frac{k_{y}^2}{\omega^2}\frac{e\varphi}{m_i}.
\end{equation}

The basic equations for electrons are the mass and momentum conservation equations with electron gyro-viscosity, electron inertia and collisions taken into consideration:

\begin{equation}
     \frac{\partial n_e}{\partial t} +  \triangledown \cdot (n_e \mathbf{v_e}) = 0,
\end{equation}

\begin{equation}
     n_em_e\frac{\partial \mathbf{v_e}}{\partial t} +  n_em_e(\mathbf{v_e} \cdot \triangledown)  \mathbf{v_e} = en_e(-\triangledown \varphi+\mathbf{v_e}\times \mathbf{B})-\triangledown p_e-\triangledown \cdot \Pi - m_en_e\nu_{en}\mathbf{v_e}
\end{equation}

Solution of Eq. A6 and Eq. A7 gives the electron density perturbation:

\begin{equation}
\frac{\widetilde{n_e}}{n_{e0}}=\frac{\omega_d+k_{y}^2\rho_e^2(\omega-\omega_E+i\nu_{en})}{\omega-\omega_E+k_{y}^2\rho_e^2(\omega-\omega_E+i\nu_{en})}\frac{e\varphi}{T_e},
\end{equation}

\noindent where $n_{e0}$ is the equilibrium electron density. The Poisson equation relates the fluctuations of electron density and ion density:

\begin{equation}
    \triangledown^2{\varphi}=\frac{e}{\varepsilon_0}(\widetilde{n_e}-\widetilde{n_i}),
\end{equation}

Inserting Eq. A5 and Eq. A8 to Eq. A9, we have:

\begin{equation}
\triangledown^2{\varphi}=\frac{en_{e0}}{\varepsilon_0}(\frac{\omega_d+k_{y}^2\rho_e^2(\omega-\omega_E+i\nu_{en})}{\omega-\omega_E+k_{y}^2\rho_e^2(\omega-\omega_E+i\nu_{en})}\frac{e\varphi}{T_e}-\frac{\alpha k_{y}^2}{\omega^2}\frac{e\varphi}{m_i}),
\end{equation}

\noindent where $\alpha=n_{i0}/n_{e0}$. The linearization of the Laplace operator gives the following expression:

\begin{equation}
k_{y}^2\varphi=\frac{en_{e0}}{\varepsilon_0}( \frac{\alpha k_{y}^2}{\omega^2}\frac{e\varphi}{m_i}- \frac{\omega_d+k_{y}^2\rho_e^2(\omega-\omega_E+i\nu_{en})}{\omega-\omega_E+k_{y}^2\rho_e^2(\omega-\omega_E+i\nu_{en})}\frac{e\varphi}{T_e}),
\end{equation}

After the cancellation and the simplification, we get Eq. 4 in the main text:

\begin{equation}
(k_{y}\lambda_{De})^2=\frac{\alpha k^2_{y}c^2_s}{\omega^2} - \frac{\omega_{d}+k^2_{y}\rho^2_{e}(\omega-\omega_{E}+i\nu_{en})}{\omega-\omega_{E}+k^2_{y}\rho^2_{e}(\omega-\omega_{E}+i\nu_{en})}.
\end{equation}

\section*{References}
%\begin{thebibliography}{}
%  \input{main.bbl}
%\end{thebibliography}
%\section*{References}
\bibliography{main.bbl}

%\bibliography{reference}

\end{document}